\documentclass[preprint,12pt]{elsarticle}

\usepackage{graphicx}%
\usepackage{multirow}%
\usepackage{amsmath,amssymb,amsfonts}%
\usepackage{amsthm}%
\usepackage{mathrsfs}%
\usepackage[title]{appendix}%
\usepackage{xcolor}%
\usepackage{textcomp}%
\usepackage{manyfoot}%
\usepackage{booktabs}%
\usepackage{algorithm}%
\usepackage{algorithmicx}%
\usepackage{algpseudocode}%
\usepackage{listings}%
\usepackage{hyperref}

\newcommand{\mgbt}{\texorpdfstring{$\mathrm{Mn}_{1-x}\mathrm{Ge}_x\mathrm{Bi}_2\mathrm{Te}_4$}}
\newcommand{\mgbthalf}{$\mathrm{Mn}_{0.5}\mathrm{Ge}_{0.5}\mathrm{Bi}_2\mathrm{Te}_4$}
\newcommand{\mgbtthird}{$\mathrm{Mn}_{0.67}\mathrm{Ge}_{0.33}\mathrm{Bi}_2\mathrm{Te}_4$}
\newcommand{\mgbtthreeeights}{$\mathrm{Mn}_{0.625}\mathrm{Ge}_{0.375}\mathrm{Bi}_2\mathrm{Te}_4$}
\newcommand{\mabt}{$\mathrm{Mn}_{1-x}\mathrm{A}_x\mathrm{Bi}_2\mathrm{Te}_4$}

\newcommand{\fgbtthreeeights}{$\mathrm{Fe}_{0.625}\mathrm{Ge}_{0.375}\mathrm{Bi}_2\mathrm{Te}_4$}
\newcommand{\mgbtthreeeightstese}{$\mathrm{Mn}_{0.625}\mathrm{Ge}_{0.375}\mathrm{Bi}_2\mathrm{Te}_{4-x}\mathrm{Se}_x$}
\newcommand{\fgbtthreeeightstese}{$\mathrm{Fe}_{0.625}\mathrm{Ge}_{0.375}\mathrm{Bi}_2\mathrm{Te}_{4-x}\mathrm{Se}_x$}
\newcommand{\fgbts}{$\mathrm{Fe}_{1-x}\mathrm{Ge}_x\mathrm{Bi}_2\mathrm{Te}_{4-y}\mathrm{Se}_y$}

\newcommand{\mbt}{$\mathrm{Mn}\mathrm{Bi}_2\mathrm{Te}_4$}
\newcommand{\fbt}{$\mathrm{Fe}\mathrm{Bi}_2\mathrm{Te}_4$}
\newcommand{\mbs}{$\mathrm{Mn}\mathrm{Bi}_2\mathrm{Se}_4$}
\newcommand{\mst}{$\mathrm{Mn}\mathrm{Sb}_2\mathrm{Te}_4$}
\newcommand{\bite}{$\mathrm{Bi}_2\mathrm{Te}_3$}
\newcommand{\gz}{$\Gamma Z$}
\newcommand{\zg}{$Z' \Gamma$}
\newcommand{\gk}{$\Gamma K$}
\newcommand{\zgz}{$Z' \Gamma Z$}
\newcommand{\kgz}{$K \Gamma Z$}

\newcommand{\lsoc}{\lambda_\text{SOC}}
\newcommand{\pconfig}{P-con\-fi\-gu\-ra\-ti\-on}
\newcommand{\xconfig}{X-con\-fi\-gu\-ra\-ti\-on}
\newcommand{\mub}{\mu_\text{B}}

\journal{Journal of Physics and Chemistry of Solids}

\begin{document}

\begin{frontmatter}



\title{Topological phase control in \mgbt{} via spin-orbit coupling and magnetic configuration engineering}

\author[spbu]{A.M. Shikin\corref{cor1}}
\ead{ashikin@inbox.ru}

\author[spbu,ufa]{N.L. Zaitsev}

\author[spbu]{A.V. Eryzhenkov}

\author[spbu]{R.V. Makeev}

\author[spbu]{T.P. Estyunina}

\author[spbu]{D.A. Estyunin}

\author[spbu]{A.V. Tarasov}

\cortext[cor1]{Corresponding author}

\affiliation[spbu]{%
  organization={Saint Petersburg State University},%
  addressline={},%
  city={Saint Petersburg},%
  postcode={198504},%
  country={Russia}}

\affiliation[ufa]{%
  organization={Institute of Molecule and Crystal Physics, Subdivision of the Ufa Federal Research Centre of the Russian Academy of Sciences},%
  addressline={},%
  city={Ufa},%
  postcode={450075},%
  country={Russia}}

\begin{abstract}
Magnetic topological systems based on \mbt{} have recently attracted significant attention due to their rich interplay between magnetism and topological electronic states. In this work, using density functional theory (DFT), we investigate topological phase transitions (TPTs) in \mgbt{} compounds with both ferromagnetic (FM) and antiferromagnetic (AFM) ordering under variations of spin-orbit coupling (SOC) strength and uniaxial strain along the $c$ axis. We show that the emergence of a Weyl semimetal (WSM) phase requires the crossing of bands with opposite $s_z$ spin projections along the \gz{} direction. Modulation of SOC and strain can annihilate Weyl points via spin-selective hybridization, driving transitions into trivial or topological insulating phases. 
Furthermore, we demonstrate that local asymmetry in Mn/Ge substitution, particularly at 37.5\% Ge concentration (\mgbtthreeeights{}) can locally disrupt AFM interlayer coupling and induce a WSM state even in globally AFM systems, without external remagnetization.
To optimize Weyl point separation and enhance the anomalous Hall effect (AHE), we propose partial substitution of Mn by Fe and Te by Se.
\end{abstract}




\end{frontmatter}



\section{Introduction}\label{sec1}

The family of magnetically ordered topological insulators (TIs) based on \mbt{} has recently attracted significant attention due to their unique combination of magnetic and topological properties, as well as the potential for controlled modification of their electronic and spin structure~\cite{tokura2019magnetic, liu2020robustaxion, wang2021intrinsic, li2024progressmbt, otrokov2019prediction, li2019intrinsic, zhang2019topological, li2019controllable, gong2019experimental, chen2019topological, estyunin2020signatures, shikin2020nature, shikin2023routes, shikin2021sample}. This material is
        characterized by an FM type of interaction within each
        Te-Bi-Te-Mn-Te-Bi-Te septuple layer (SL) block and an AFM
        interaction between neighboring SLs, forming the bulk of the
        crystal. Due to the ordered arrangement of magnetic atoms in the crystal
        structure and their high concentration, as well as the enhanced
        SOC induced by the influence of heavy atoms, this
        type of TI exhibits a strong interplay between topology and magnetism.
        This enables the realization of various quantum topological effects -- most notably the quantum anomalous Hall effect (QAHE) -- at significantly higher temperatures than those achievable in magnetically doped TIs~\cite{tokura2019magnetic, liu2020robustaxion, wang2021intrinsic, li2024progressmbt, deng2020quantum}. For example, QAHE has been observed in thin \mbt{} films at 1.4~K in zero magnetic field and up to 6.5~K under an applied magnetic field~\cite{deng2020quantum}.

        Furthermore, replacing Mn atoms with nonmagnetic elements Ge(Sn,Pb)
        leads to a decrease in the bulk band gap in \mabt{} (A = Ge, Sn, Pb)
        compounds, nearly reaching zero at substitution concentrations of
        40--50\% \cite{qian2022magnetic, changdar2023nonmagnetic,
        zhu2021magnetic, tarasov2023msbt, estyunina2023evolution,
        frolov2024magnetic, shikin2025mgbt, estyunin2024mpbt}, with a presumed
        transition (for the FM phase) from a topological insulator to a Weyl
        semimetal (WSM) state \cite{shikin2025mgbt}. At even higher substitution levels (above 80\%), a reentrant transition to a trivial or weakly topological insulating phase is typically observed~\cite{qian2022magnetic, changdar2023nonmagnetic, zhu2021magnetic, tarasov2023msbt, estyunina2023evolution, frolov2024magnetic, shikin2025mgbt, estyunin2024mpbt}.  
        
        Additionally, increasing the concentration of nonmagnetic dopants significantly modifies the magnetic properties of AFM \mabt{} compounds. Specifically, dilution of magnetic Mn atoms lowers the Néel temperature and reduces the spin-flop transition field~\cite{qian2022magnetic,
        changdar2023nonmagnetic, zhu2021magnetic, estyunin2023comparative}, thereby promoting conditions favorable for the formation of magnetic WSM phases and enabling further exploration of their physicochemical behavior.

        The growing interest in WSMs stems from their unique properties,
        including the anomalous spin Hall effect, chiral anomaly effect, chiral
        magnetic effect, and negative magnetoresistance effect (see, for
        example, reviews \cite{armitage2018weyl, lv2021semimetals,
        hasan2021highfoldweyl, yan2017weyl, shen2017weyl, kar2021weylprimer}).
        At the same time, for WSMs the above-mentioned effects are of a bulk
        rather than surface character, which significantly increases the
        efficiency of their practical use.
        Unlike magnetic TIs, magnetic WSMs exhibit gapless crossings of state
        branches with opposite spin orientations while preserving their
        nontrivial topological properties~\cite{shen2017weyl,
        armitage2018weyl, lv2021semimetals, hasan2021highfoldweyl, yan2017weyl,
        kar2021weylprimer}. A
        recent study \cite{belopolski2025weylferromagnet} experimentally
        demonstrated the realization of a WSM phase in Cr-doped TI \bite{},
        featuring an enhanced anomalous Hall effect, negative magnetoresistance,
        and other unique effects inherent to WSMs. This finding opens the door to broader application of TI-based magnetic WSMs in modern spintronics.

        In light of the above, the search for materials with bulk magnetic WSM
        properties, as well as the analysis of their physicochemical properties
        and formation conditions, represents an important and actual scientific
        and technical challenge. The Weyl nodes formed in such systems -- carrying opposite chirality -- act as sources of Berry curvature in momentum space and are often viewed as analogs of magnetic monopoles and antimonopoles in k-space~\cite{lv2021semimetals, hasan2021highfoldweyl, yan2017weyl, shen2017weyl}, further fueling interest in the fundamental and applied study of WSMs.

        In condensed matter physics, the works analyzing WSM formation and the
        features of their electronic structure, as well as transition conditions
        to the WSM state (see, for example,
        \cite{wang2021intrinsic, li2019intrinsic, zhang2019topological,
        li2019controllable, armitage2018weyl, lv2021semimetals,
        hasan2021highfoldweyl, yan2017weyl, shen2017weyl, kar2021weylprimer,
        wang2021weylstabilize, chowdhury2019prediction, zhang2021tunable,
        zhou2020topological, li2021mbt, guo2023novelfbt}), have shown that such
        a transition can occur due to the breaking of either spatial symmetry or
        time-reversal symmetry (TRS). Moreover, the works \cite{wang2021intrinsic,
        li2019intrinsic, zhang2019topological, li2019controllable,
        shikin2025mgbt, wang2021weylstabilize, chowdhury2019prediction,
        zhang2021tunable, zhou2020topological, li2021mbt, guo2023novelfbt}
        demonstrated that WSM formation in magnetic TIs of the \mbt{} family
        (including \mst{}, \mbs{}, etc.) occurs precisely due to TRS breaking under 
        FM ordering which determines the possibility of gapless crossings of state branches with
        opposite spin orientations \cite{armitage2018weyl, lv2021semimetals,
        hasan2021highfoldweyl, yan2017weyl, shen2017weyl, kar2021weylprimer}.
        For this type of material, such crossings occur in the \gz{}
        direction of the Brillouin zone (BZ), forming two symmetric Weyl points (WPs) of
        opposite chirality in each $-Z\Gamma$ and \gz{} directions~\cite{wang2021intrinsic,
        li2019intrinsic, zhang2019topological, li2019controllable,
        wang2021weylstabilize, chowdhury2019prediction, zhang2021tunable,
        zhou2020topological, li2021mbt}.

        In this work, we conduct a comparative theoretical analysis of TPTs in \mgbt{} with both AFM and FM interactions, focusing on the corresponding changes in electronic structure and the underlying factors governing these transitions. We present band structure calculations across varying Ge concentrations in the AFM and FM phases of \mgbt{}, and examine possible TPTs between the magnetic TI and normal insulator (NI) phase. Particular attention is paid to intermediate Weyl and Dirac semimetal states arising under different parameter variations, enabling identification of the key conditions for the realization of each phase.

        In the first part, we analyze the evolution of the electronic structure of \mgbt{} with AFM and FM interactions as a function of Ge concentration (via Mn~\(\rightarrow\)~Ge substitution), aiming to identify concentration ranges with minimal bulk band gap -- a necessary condition for Weyl phase formation. To further determine the conditions under which the Weyl phase emerges and to examine related TPTs, we study how the bulk electronic structure responds to variations in spin-orbit coupling strength (\(\lambda_{\mathrm{SOC}}\)) and uniaxial strain \(\gamma_c\) along the crystallographic \(c\)-axis, for both AFM and FM configurations in systems with minimal band gaps. This includes analysis of the Weyl phase stability boundaries and the regions of TPTs between WSM, TI, and NI phases, as revealed by inversion patterns of Te \(p_z\) and Bi \(p_z\) orbital contributions at the band edges. These TPTs are intimately linked to the spin structure of the relevant bands, which is also investigated.

In the last part, we explore the possibility of Weyl phase formation in a system with an initially AFM interaction (without requiring external remagnetization), taking into account possible Mn/Ge substitution configurations that disrupt local AFM interlayer interactions. Additionally, we assess the modulation of Weyl phase parameters by varying the effective magnetic moment, replacing Mn with another magnetic metal (Fe), and introducing Te/Se substitutions.

\section{Methods}

        First-principles calculations in the framework of the density functional theory
(DFT) were partially performed at the Computing Center of SPbU Research park. The
electronic structure calculations with impurities were conducted using
the OpenMX DFT code, which implements a linear
        combination of pseudo-atomic orbitals (LCPAO) approach
        \cite{ozaki2003variationally, ozaki2004numerical, ozaki2005efficient}
        with full-relativistic norm-conserving pseudopotentials
        \cite{troullier1991efficient}. Exchange-correlation functional in the
        GGA-PBE form was employed \cite{perdew1996generalized} and basis sets
        were specified as Ge7.0-s3p2d2, Mn6.0-s3p2d1, Fe6.0S-s3p2d1,
        Te7.0-s3p2d2f1 and Bi8.0-s3p2d2f1. Here the number following the
        chemical species represents the range of the basis set in Bohr radii and its
        size is given by the number of primitive functions for each channel.
        Real-space numerical integration accuracy was specified by a cutoff
        energy of 200 Ry, total-energy convergence criterion was set to
        $10^{-6}$~Hartree.  Localized Mn~$3d$ and Fe~$3d$
        states were considered using DFT + $U$ \cite{hanozaki2006ldau} within
        Dudarev approach \cite{dudarev1998cj} with $U = 5.4$~eV and $U =
        4.5$~eV, respectively.  Brillouin zone was sampled using $7 \times 7
        \times 3$ grid.
        
        Simulations of bulk \mgbt{} compounds for $x = 25\%$, $x = 37.5\%$ and
        $x = 50\%$  were performed using $2 \times 2$ supercells ($5 \times 5
        \times 3$ BZ sampling), $x = 33\%$ case was studied using $3 \times 1$
        supercell ($5 \times 7 \times 3$ BZ sampling) and cases of $x = 40\%$
        and $x = 60\%$ were calculated using $5 \times 1$ supercells ($3 \times
        7 \times 3$ BZ sampling). Unit cells in all calculations included 2~SLs.

    \section{Results and discussion}
    \subsection{Bulk band structure of \mgbt{} in AFM and FM phases}

        We begin with analysis of zero band gap conditions in AFM and FM phases
        of  \mgbt{} which is necessary for the formation of the WSM phase
        through gradual substitution of magnetic Mn atoms with non-magnetic Ge
        atoms (increasing $x$). We also consider possible TPTs that may occur in
        the process. 

Two different types of Mn/Ge substitution site configurations in neighboring SLs are analyzed. If both Mn layers in adjacent SLs are doped at identical positions, the configuration is referred to as “parallel” (\pconfig). If the Mn/Ge substitution sites differ between neighboring SLs, the configuration is termed “cross” (\xconfig). These two configurations represent limiting cases; real systems are likely to exhibit intermediate arrangements. Band structures for both \pconfig{} and \xconfig{} geometries are considered in this work, with schematic representations shown in Fig.~1S (Suppl.).
        
\begin{figure}[ht!]
            \includegraphics[width=13.5cm]{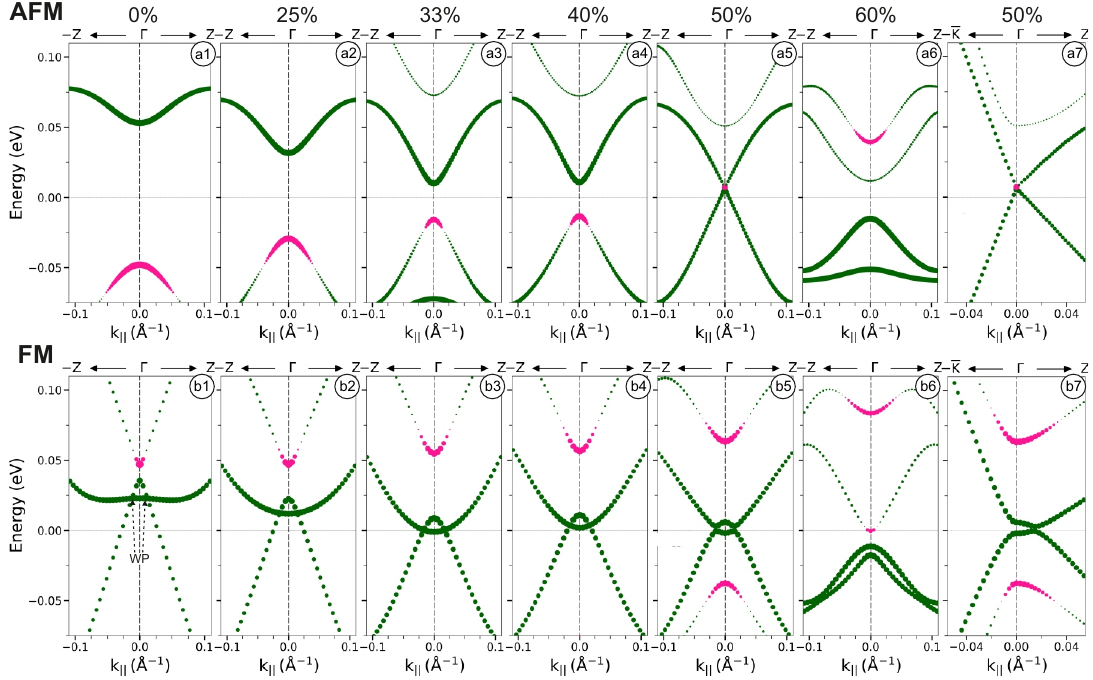}

            \caption{Bulk band structure calculations along the \zgz{}
            direction of the BZ for \mgbt{} in the \pconfig{} for different Ge concentrations
            between 0\% and 60\% for AFM (a1--a6) and FM (b1--b6) phases. 
            States where Te~$p_z$ or Bi~$p_z$ contributions
            are dominant are indicated by green and pink colors, respectively.
            Panels (a7) and (b7) show band structures along the \kgz{} direction
            for \mgbthalf{} which attains the minimum possible bulk band gap in the \pconfig{}.}
            \label{fig1}
        \end{figure}

        Figures~\ref{fig1}(a1--a6) present comparative changes in the bulk
        electronic structure for the AFM phase of \mgbt{} during Mn/Ge
        substitution for configurations with equivalent substitution sites in
        neighboring SLs (\pconfig), shifted by half a unit cell. This
        arrangement is characterized by minimal disruption of the AFM
        interaction between neighboring SLs. At the same time,
        Figs.~2S(a1--a5) Suppl. show the changes in electronic
        structure for an asymmetric order of the Mn/Ge substitution sites in
        neighboring SLs (\xconfig). This configuration is characterized by some
        local disruption of the AFM order due to the increased contribution of
        local FM-like interactions between Mn layers mediated through the Ge
        $sp$ states~\cite{yang2025septuple}. The presented band dispersions
        generally correlate (where parameter ranges overlap) with those
        calculated along the \kgz{} direction of the Brillouin zone for \mabt{}
        systems (where A = Ge, Sn, Pb) in Refs.~\cite{tarasov2023msbt,
        estyunina2023evolution, frolov2024magnetic, shikin2025mgbt,
        estyunin2024mpbt, tarasov2025prb}.
        
        Figures~\ref{fig1}(b1--b6) present the corresponding changes in the bulk
        electronic structure for the FM phase in \pconfig{}. Similar changes in
        the bulk electronic structure for \xconfig{} are shown in
        Fig.~2S(b1--b5) Suppl. It can be seen that for both types of
        configurations the original \mbt{} system with FM coupling between
        neighboring SLs already possesses a bulk electronic structure
        characteristic of a WSM phase. Across all presented dispersions the
        regions with dominant Bi $p_z$ and Te $p_z$ contributions are marked
        with pink and green symbols, respectively. 
        The band structure were calculated assuming an out-of-plane orientation of Mn magnetic moments (i.e., perpendicular to the basal \(ab\)-plane). Mn atoms are FM-coupled within each SL, and are AFM- or FM-coupled between neighboring SLs in the AFM and FM phases, respectively~\cite{otrokov2019prediction, li2019intrinsic, zhang2019topological, li2019controllable, gong2019experimental, chen2019topological, estyunin2020signatures, shikin2020nature, shikin2023routes}. The dispersions are shown along the \zgz{} direction of the BZ, where Weyl points are located in this system, making the WSM-related features most apparent.

For comparison, Figs.~\ref{fig1}(a7,~b7) and Figs.~2S(a6,~b6) (Suppl.) show band dispersions calculated along the \kgz{} direction for systems with minimal bulk band gap values (50\% and 33\% Ge concentrations for \pconfig{} and \xconfig{}, respectively), allowing one to track the continuation of relevant dispersion branches in the \gk{} direction. Formation of Weyl points with opposite chirality is only possible when the valence band (VB) and conduction band (CB) branches intersect.

The analysis shows that in the AFM phase, for both Mn/Ge substitution configurations, increasing Ge concentration leads to a progressive reduction of the bulk band gap at the \(\Gamma\)-point, reaching near-zero values. However, the critical Ge concentration depends on the configuration: approximately 50\% for \pconfig{} and 33\% for \xconfig{}. With further Ge doping, the band gap reopens. At the point of minimal gap, a gapless, cone-like linear dispersion forms, characteristic of a Dirac semimetal (DSM). In this regime, inversion of Te \(p_z\) and Bi \(p_z\) states occurs at the Dirac point (DP), where VB and CB edges exhibit opposite parity. This reflects a TPT from a state with inverted Te/Bi \(p_z\) orbital character at the \(\Gamma\)-point, typical of a TI~\cite{shikin2023routes, tarasov2023msbt, shikin2025mgbt, shikin2023topological, shikin2024study}, to a non-inverted configuration characteristic of a NI.

In the FM phase, the transition into the WSM state is more clearly manifested and is already present in pristine \mbt{} for both Mn/Ge substitution configurations. The emergence of the WSM phase in \mbt{} with FM-type interlayer coupling has also been noted in Refs.~\cite{li2019intrinsic, zhang2019topological, li2019controllable}. This phase is marked by gapless crossings of Te \(p_z\)--Bi \(p_z\) states near the DP (with a predominant Te \(p_z\) contribution, as shown in green at the \(\Gamma\)-point), which is a defining feature of WSM behavior~\cite{chowdhury2019prediction}.

According to Refs.~\cite{li2019controllable, chowdhury2019prediction, guo2023novelfbt}, the WSM phase in FM systems arises due to Zeeman-induced band splitting, which leads to the formation of gapless Weyl points (WPs in Fig.~\ref{fig1}) located symmetrically along the \zgz{} direction. These points possess opposite chiralities and are topologically protected~\cite{armitage2018weyl, lv2021semimetals, hasan2021highfoldweyl, yan2017weyl, shen2017weyl, kar2021weylprimer}. 

External perturbations in the WSM state may shift the Weyl points along the \zgz{} axis but cannot open a gap at their location, preserving their gapless nature until the WPs eventually merge at the \(\Gamma\)-point and annihilate~\cite{armitage2018weyl, lv2021semimetals, hasan2021highfoldweyl, yan2017weyl, shen2017weyl, kar2021weylprimer}.

The branches forming the Weyl points are surrounded by bands with dominant Bi~\(p_z\) character at the \(\Gamma\)-point (see \mbt{}~\cite{li2019controllable}, \mbs{}~\cite{chowdhury2019prediction}, and \fbt{}~\cite{guo2023novelfbt} for comparison). These states are marked with pink symbols in Figs.~\ref{fig1}(b1--b6) and Figs.~2S(b1--b5) (Suppl.).  As discussed below, annihilation of the Weyl points occurs when bands with the same spin orientation intersect, leading to the opening of a gap at the crossing point.

        The dispersions shown in Fig.~\ref{fig1}(b7) and Fig.~2S(b6)
        Suppl. along the \kgz{} direction of the BZ show the VB and CB state
        intersections forming Weyl points with opposite chirality.  They show no
        fundamental changes of bulk band structures for the FM phase for
        different Mn/Ge substitution configurations.  In the AFM phase such band
        intersections in the \gz{} directions are not observed. 

    \subsection{Bulk band structure of \mgbt{} under variation of SOC strength and $c$-uniaxial strain}

In this subsection, we provide a detailed analysis of the features and formation conditions of the WSM and DSM phases in \mgbt{} with FM and AFM ordering. We also examine the associated TPTs and the corresponding evolution of the band structure.

\begin{figure}[ht!]
            \centering\includegraphics[height=13cm]{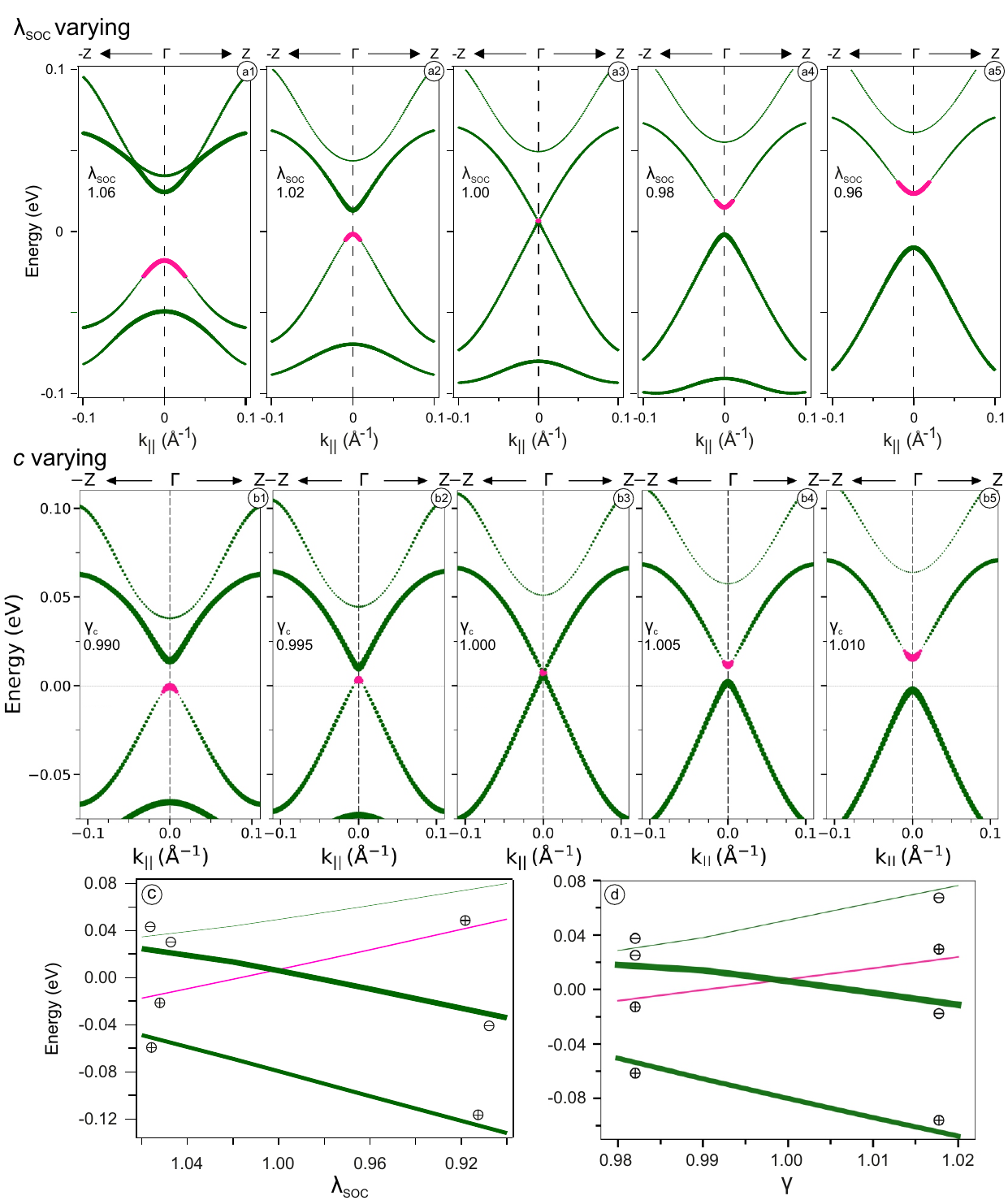}

            \caption{Bulk band structure calculations along the \zgz{} direction
            for AFM \mgbthalf{} in \pconfig{} with (a1--a5) $\lsoc$ modulation and (b1--b5) $\gamma_c$
            modulation ($\lsoc = \gamma_c = 1$ correspond to Fig.~\ref{fig1}).
            States where Te~$p_z$ or Bi~$p_z$ contributions are dominant are
            indicated by green and pink colors, respectively. Energy
            level and parity diagrams for $\Gamma$-point eigenstates with
            dominant Te~$p_z$ (green) or Bi~$p_z$ (pink) contributions for
            (c) $\lsoc$ and (d) $\gamma_c$ variation, respectively.  Line thicknesses
            correspond to magnitude of $|\text{Te } p_z - \text{Bi } p_z|$
            difference.}
    
            \label{fig2}
        \end{figure}

\paragraph{AFM phase.} Fig.~\ref{fig2}(a1--a5) and Fig.~3S(a1--a5) (Suppl.) present the calculated bulk band structures for AFM \mgbthalf{} with \pconfig{} and AFM \mgbtthird{} with \xconfig{}, respectively, along the \zgz{} direction of the BZ, where WSM-like features are most clearly manifested. These calculations were performed by varying \(\lambda_{\mathrm{SOC}}\) for systems exhibiting the smallest bulk band gap at the \(\Gamma\)-point for \(x = 50\%\) (\pconfig{}) and \(x = 33\%\) (\xconfig{}), based on the results shown in Fig.~\ref{fig1} and Fig.~2S (Suppl.). The value \(\lambda_{\mathrm{SOC}} = 1\) corresponds to the normal SOC strength used in those reference figures. The color scale indicating Te/Bi \(p_z\) contributions is consistent with that in Fig.~\ref{fig1} and Fig.~2S (Suppl.).

Additionally, Figs.~\ref{fig2}(b1--b5) and Figs.~3S(b1--b5) (Suppl.) show the evolution of the bulk band structure for the respective systems (\pconfig{} at \(x = 50\%\) and \xconfig{} at \(x = 33\%\)) under varying uniaxial strain \(\gamma_c\), where \(\gamma_c = 1\) corresponds to the unstrained case shown in Fig.~\ref{fig1} and Fig.~2S (Suppl.). Application of such uniaxial strain leads to simultaneous modification of both exchange and spin-orbit interactions to some extent; see ref.~\cite{guo2023novelfbt}.

        Fig.~\ref{fig2}(c) shows the energy and parity diagram for the $\Gamma$-point
        for four bands closest to the initial DP position where Te~$p_z$ and Bi~$p_z$ 
        contributions are also depicted. The parity was determined by the expression 
        $
            P = \operatorname{sign} \int \limits_\text{unit cell} \psi^*(\mathbf r) \psi(-\mathbf r)\: \mathrm d\mathbf r
        $
        where $\psi$ is the $\Gamma$-point eigenstate wave function in question.

        As shown by the presented dispersions, for the AFM phase for both Mn/Ge
        substitution configurations, variation of $\lsoc$
        (Fig.~\ref{fig2}(a1--a5) and Fig.~3S(a1--a5) Suppl.) leads to a
        TPT from TI into NI phase, accompanied by both parity and Te/Bi~$p_z$
        band inversion at the DP (see also~\cite{shikin2023routes,
        shikin2025mgbt, shikin2023topological, shikin2024study}). 
        
        For AFM \mgbthalf{} this TPT occurs at a single $\lsoc$ critical point
        which corresponds to an intermediate DSM stage which is also evident
        from Fig.~\ref{fig2}(c).  No additional intersections between VB and CB
        along the \zgz{} direction charactesictic of WSM phase are observed
        while $\lsoc$ is varied.

        Bulk band structure response to strain, shown in Fig.~\ref{fig2}(b1--b5,~d), exhibits a similar behavior: there exists a single critical value of \(\gamma_c\) where a TPT between the TI and NI phases occurs. This transition is identified by both parity inversion and the inversion of Te/Bi~\(p_z\) contributions at the DP, indicating the presence of an intermediate DSM phase~\cite{shikin2023routes, shikin2025mgbt, shikin2023topological, shikin2024study}. As in the case of SOC variation, no WSM-specific features (such as Weyl node formation) are observed along the \zgz{} direction.

        The system \mgbtthird{} with \xconfig{} of Mn/Ge substitution sites
        which also attains mininum band gap (Figs.~3S(a1--a5) Suppl.
        and Figs.~3S(b1--b5) Suppl.) demonstrates the same properties
        regarding the existence of only one critical TPT point either for
        $\lsoc$ or $\gamma_c$ variation where TI phase transforms into NI phase
        through the DSM phase.

\begin{figure}[H]
            \centering\includegraphics[width=13cm]{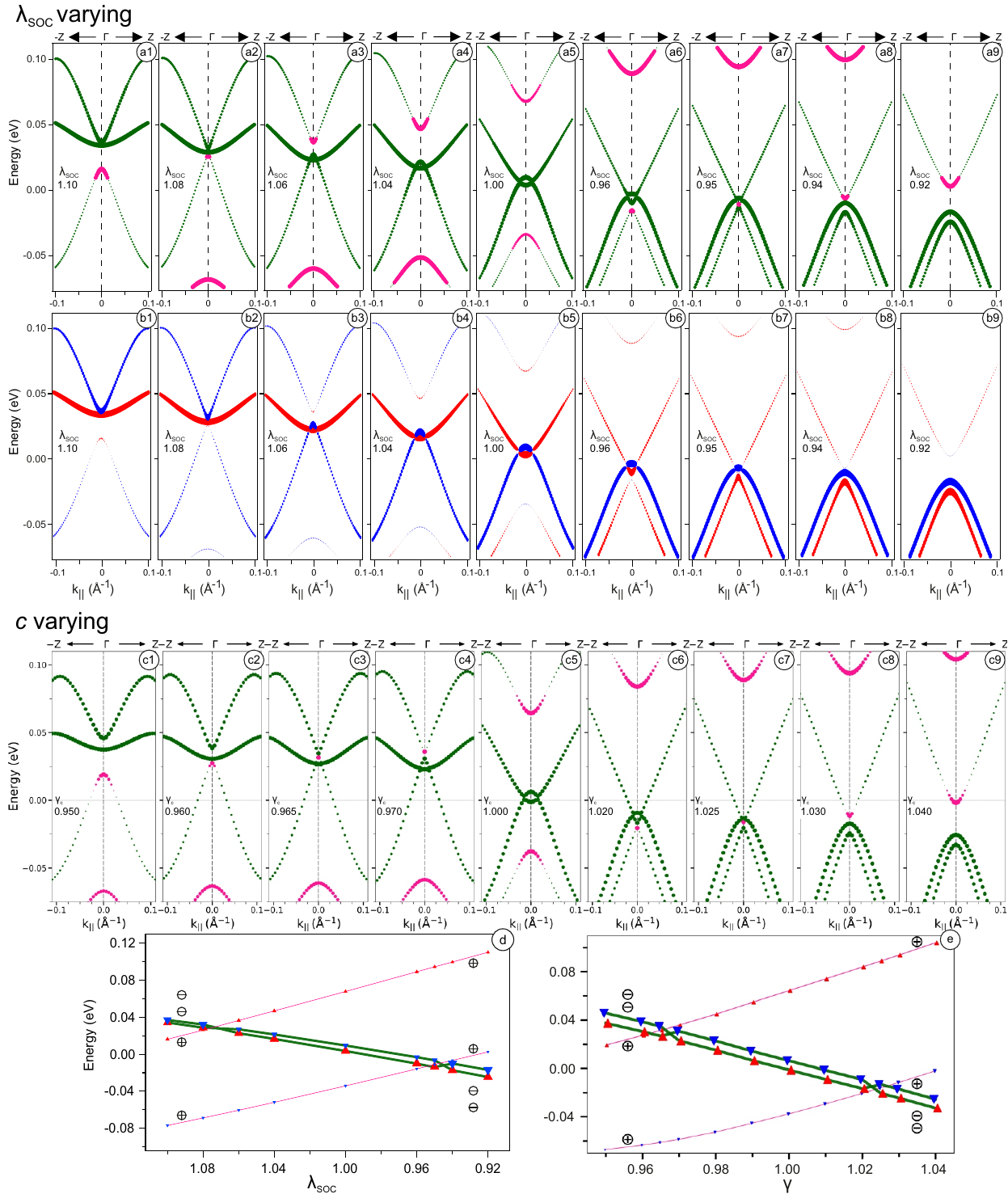}

            \caption{Band structure calculations along the \zgz{} direction for
            FM \mgbthalf{} in \pconfig{} when $\lsoc$ is varied (a1--a9),
            showing their corresponding $s_z$ ($z \perp ab\text{-plane}$) spin
            structures in (b1--b9), and when strain $\gamma_c$ is applied
            (c1--c9).  Energy level and parity diagrams for $\Gamma$-point
            eigenstates with dominant Te~$p_z$ (green) or Bi~$p_z$ (pink)
            contributions as well as opposite spin directions in red and blue
            for (d) $\lsoc$ and (e) $\gamma_c$ variation, respectively. Line
            thicknesses correspond to magnitude of $|\text{Te } p_z - \text{Bi }
            p_z|$ difference.}
    
            \label{fig3}
        \end{figure}

        \paragraph{FM phase.} Fig.~\ref{fig3}(a1--a9) and
        Fig.~4S(a1--a9) present the results of band structure
        calculations along the \zgz{} direction for FM \mgbthalf{} (\pconfig)
        and FM \mgbtthird{} (\xconfig), respectively, where $\lsoc$ was varied.
        Fig.~\ref{fig3}(c1--c9) and Fig.~4S(b1--b9) show band
        structures of FM \mgbthalf{} (\pconfig) and FM \mgbtthird{} (\xconfig),
        respectively, for different $\gamma_c$ values.

        Both initial FM-ordered phases of \mgbthalf{} (\pconfig) and
        \mgbtthird{} (\xconfig) with $\lsoc = 1$ (Fig.~\ref{fig3}(a5)) or
        $\gamma_c = 1$ (Fig.~\ref{fig3}(c5)) are already in the WSM phase which
        is signified by two explicit band intersections along the \zg{}  and
        \gz{} directions resulting in two Weyl nodes which are located
        symmetric about the $\Gamma$-point.
        
Band structures in Figs.~\ref{fig3}(a1--a9) and \ref{fig3}(c1--c9) reveal a clear qualitative similarity between the effects of crystal compression (\(\gamma_c < 1\)) and SOC enhancement (\(\lambda_{\mathrm{SOC}} > 1\)). Conversely, crystal tension (\(\gamma_c > 1\)) corresponds qualitatively to SOC reduction (\(\lambda_{\mathrm{SOC}} < 1\)). An increase of \(\lambda_{\mathrm{SOC}}\) to 1.10 (Figs.~\ref{fig3}(a1,b1)) or a decrease of \(\gamma_c\) to 0.95 (Fig.~\ref{fig3}(c1)) results in a TPT from the FM-ordered WSM phase to an FM TI phase. On the other hand, a significant reduction of \(\lambda_{\mathrm{SOC}}\) to 0.92 or an increase of \(\gamma_c\) to 1.03 leads to a transition into the NI phase. In both scenarios, two Weyl points annihilate at the \(\Gamma\)-point, giving rise to a gapped phase—with either inverted (TI) or normal (NI) ordering of Te/Bi orbital contributions at the \(\Gamma\)-point; see~\cite{shikin2023routes, tarasov2023msbt, shikin2025mgbt, shikin2023topological, shikin2024study} for comparison. These results are consistent with observations reported for \fbt{} in~\cite{guo2023novelfbt}.

        Under transition from TI state, starting from $\lsoc=1.06\ldots1.08$ and $\gamma_c=0.965$, the lowermost
        branch of CB states intersects the uppermost branch of VB states along
        the \zgz{} direction by two Weyl points, thus forming the WSM phase. At
        either $\lsoc = 1.06$ or $\gamma_c = 0.97$, the Te/Bi $p_z$ contribution
        order becomes inverted, thus completing the TPT from TI to a WSM phase.

The WSM phase with two Weyl points remains stable within the ranges \(1.06 \geqslant \lambda_{\mathrm{SOC}} \geqslant 0.96\) or \(0.97 \leqslant \gamma_c \leqslant 1.02\). In either case, the Bi \(p_z\)-dominated branches of both the VB and CB shift upward in energy, and the separation between Weyl points increases, reaching a maximum near \(\lambda_{\mathrm{SOC}} = 1\) or \(\gamma_c = 1\).

Further decrease of \(\lambda_{\mathrm{SOC}}\) to 0.96 or increase of \(\gamma_c\) to 1.02 results in noticeable changes in the band curvature of the Bi-dominated lower VB and the Te-dominated lower CB, indicating increased hybridization between these states. When either \(\lambda_{\mathrm{SOC}} < 0.95\) or \(\gamma_c > 1.025\), the Weyl points annihilate, the band structure becomes gapped, and the band ordering of these interacting branches is no longer inverted. In both cases, the system transitions into a NI phase~\cite{shikin2023routes, tarasov2023msbt, shikin2025mgbt, shikin2023topological, shikin2024study}.

        The transition from the WSM to NI phase proceeds through a Dirac cone
        stage with inversion of states with opposite parity at the DP with an
        additional intersection of the DP region by the third branch.  The
        hybridization at the DP leads to a band gap and complete destruction of
        the Weyl phase, thus defining the WSM phase boundaries.  Similar
        processes of Dirac cone formation occur also in the TPT from WSM to TI (under
        increasing $\lsoc$ and decreasing $\gamma_c$), defining the WSM
        boundaries both under increase and decrease of $\lsoc$ and
        compression/expansion of the crystal.

        Variation of $\lsoc$ and $\gamma_c$ for FM-ordered system \mgbtthird{}
        with X-configuration of Mn/Ge substitution sites
        (Fig.~4S(a,~b)) leads to similar behavior of its bulk band
        structure and features of TPTs, confirming that WSM formation is a
        general result regarding the FM-ordered \mgbt{} which does not strongly
        depend on exact positions of Mn/Ge substitution sites.

        These results correlate well with previously reported data for similar
        systems \cite{chowdhury2019prediction, zhang2019topological,
        li2019controllable, guo2023novelfbt}, also showing that Weyl points in
        FM TIs originate from crossings of states along the \zgz{} direction
        with dominant Te $p_z$ contributions \cite{chowdhury2019prediction}.

        \paragraph{Spin structure evolution.} More detailed analysis of the bulk band structure under varying \(\lambda_{\mathrm{SOC}}\) and associated TPTs in FM \mgbthalf{} (\pconfig{}) is presented in Fig.~\ref{fig3}(b1--b9), which shows the \(s_z\) (out-of-plane) spin components corresponding to the band structures in Fig.~\ref{fig3}(a1--a9). The spin textures confirm the gapless nature of the Weyl phase, consistent with previously reported results~\cite{armitage2018weyl, lv2021semimetals, hasan2021highfoldweyl, yan2017weyl, shen2017weyl, kar2021weylprimer}.

Variation of \(\gamma_c\) in the 50\% Ge system produces qualitatively similar changes in the spin structure. Fig.~\ref{fig3}(d,~e) summarizes the evolution of the energy positions of Te~\(p_z\) and Bi~\(p_z\) states at the \(\Gamma\)-point under changing \(\lambda_{\mathrm{SOC}}\) and \(\gamma_c\), including their spin orientations and parities (\(+\), \( - \)) at the \(\Gamma\)-point.

First and foremost, the spin-resolved dispersions in Fig.~\ref{fig3}(b1--b9) confirm that the FM \mgbthalf{} system with \(\lambda_{\mathrm{SOC}} = 1\) realizes a WSM phase. This phase is marked by gapless band crossings between oppositely spin-polarized  branches along the \zgz{} direction, symmetrically positioned about the \(\Gamma\)-point.

In this configuration, the Te~\(p_z\)-dominated states that form the Weyl points exhibit opposite spin polarizations, which is essential for WSM phase stabilization. Additionally, the Bi~\(p_z\)-dominated VB and CB states at the \(\Gamma\)-point also exhibit opposite spin orientations, inverted with respect to those of the Te~\(p_z\)-derived states forming the Weyl nodes.

        With a slight increase in SOC strength ($\lsoc = 1.06$) (panel (b3)),
        higher-energy branches shift downward, approaching the branches forming
        the Weyl nodes. At $\lsoc = 1.08$ an inversion of spin polarization
        occurs at the $\Gamma$-point, thereby violating the conditions for the
        gapless intersection of branches of states with opposite spin
        orientations, that leads to annihilation of the Weyl points and
        destruction of the WSM phase. At $\lsoc = 1.10$ the system becomes
        finally an FM TI (panel (b1)). In this regime clear spin splitting is
        observed for conduction band states dominated by Te~$p_z$ near the
        $\Gamma$ point. VB states with dominant Bi~$p_z$ character also exhibit
        spin splitting, with an inverted spin orientation relative to that of
        the CB, and modulation of $s_z$ spin polarization away from the
        $\Gamma$-point.

When \(\lambda_{\mathrm{SOC}}\) is reduced (panel (b6)), the system undergoes a reversal of the previously observed behavior. The Te~\(p_z\)-dominated states forming the Weyl points are approached from below by a cone of Bi~\(p_z\)-dominated states at the \(\Gamma\)-point, characterized by \((+)\) parity and spin orientation similar to that of the upper Weyl-forming branch (see Fig.~\ref{fig3}(e)).

At \(\lambda_{\mathrm{SOC}} = 0.95\), a spin inversion occurs within these Bi-derived cone states at the \(\Gamma\)-point. Because these states now share the same spin orientation as those forming the Weyl crossings, hybridization becomes allowed, leading to interaction and annihilation of the Weyl points (panel (b7)), accompanied by the opening of a gap.

As \(\lambda_{\mathrm{SOC}}\) decreases further to 0.94 and below (panel (b8)), the system enters a trivial FM insulator phase, characterized by pronounced spin splitting in the valence band. The conduction band shows a more intricate spin structure, but spin polarization remains visible at the \(\Gamma\)-point. Notably, the resulting spin structure is inverted relative to that of the FM TI phase (panel (b1)), consistent with a TPT from a TI to an NI phase.

        From the above, we conclude that in the FM phase, it is specifically the
        crossings between oppositely spin-polarized branches along \zg{} and
        \gz{} directions that lead to the formation of the Weyl phase with
        topologically protected Weyl nodes located symmetrically along these
        directions. Weyl node annihilation occurs already at the repeated
        intersection of one of the branches of states forming Weyl points with
        other branches of Te~$p_z$/Bi~$p_z$ states shifting in energy, with the
        same spin orientation.  In this case, the branches with identical spin
        orientation can hybridize, leading to the formation of an energy gap at
        the crossing point. This breaks the chiral symmetry at the Weyl nodes,
        destroys the WSM phase, and results in a transition to a NI.
        
    \subsection{Analysis of Weyl phase formation in a system with initial AFM ordering}

        A standard way to replace the AFM-type interaction with FM (in order to
        transition to the WSM phase) in systems based on AFM \mbt{} (including
        by substituting Mn atoms with Ge, Sn, or Pb) is remagnetization by an
        external magnetic field~\cite{li2019intrinsic, zhang2019topological,
        li2019controllable, shikin2025mgbt, chowdhury2019prediction,
        guo2023novelfbt, estyunin2023comparative}. However, taking into account
        the fact that when Mn atoms are partially  substituted by Ge (as well as
        Sn, Pb), the Ge $p_z$-orbitals mediate some of the Mn---Mn interaction
        between neighboring Mn layers and create  a channel for local FM
        coupling between interlayer Mn atoms~\cite{yang2025septuple,
        shikin2025mgbt}.  It may be assumed that  increase in Mn/Ge substitution
        concentration leads to accumulation of such local interlayer FM
        interactions which implies the possibility of natural formation of WSM
        phase in \mgbt{} with no external remagnetization  into an FM-ordered
        state. 

        One of the ways of breaking the AFM interlayer coupling in this type of
        TI can be  formation of asymmetric (uncompensated) local Mn/Ge
        substitutions across the whole bulk within  neighboring SLs. A more
        idealized option is alternating undoped and Mn/Ge-substituted layers in
        a pattern similar to given in \cite{burkov2014aheweyl}.
        
        \begin{figure}[H]
            \centering\includegraphics[width=\linewidth]{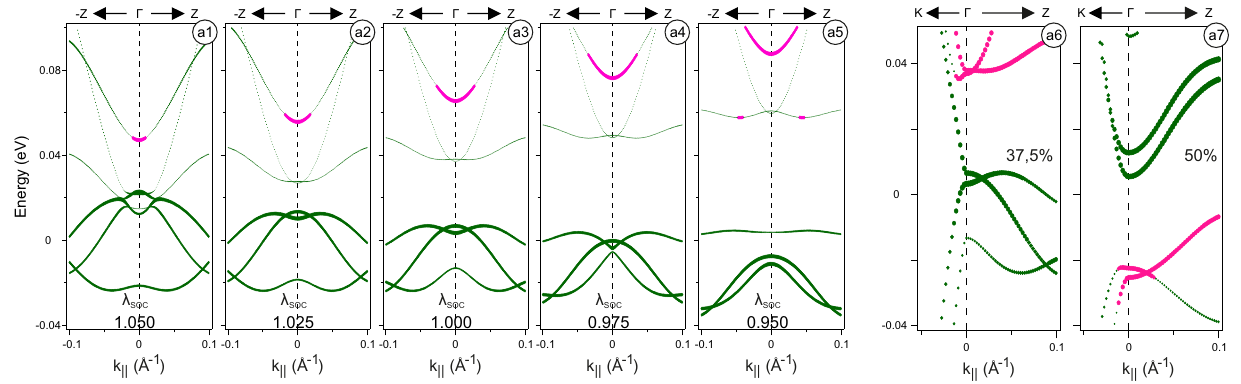}

            \caption{(a1--a5) Band structure calculations for AFM \mgbtthreeeights{} with $\lsoc$ variation along the \zgz{} direction where all Mn/Ge substitution sites are located in the same Mn layer. (a6,~a7) Corresponding band structures of AFM \mgbtthreeeights{} and AFM \mgbthalf{}, respectively, along the \kgz{} direction. States where Te~$p_z$ or Bi~$p_z$ contributions are dominant are indicated by green and pink colors, respectively.}

            \label{fig4a}
        \end{figure}

Figures~\ref{fig4a}(a1--a5) present theoretical bulk band structure calculations for AFM \mgbt{} along the \zgz{} direction at varying \(\lambda_{\mathrm{SOC}}\), for \(x = 37.5\%\) Ge concentration—close to the composition where the bulk band gap reaches a minimum. Green and pink symbols indicate dominant Te~\(p_z\) and Bi~\(p_z\) orbital contributions, respectively.

Some calculations of the bulk band structure along the \kgz{} direction for related systems with alternating Mn/Ge substitution patterns were previously reported in~\cite{shikin2025mgbt}.

This system is analogous to the layered structures of the \mbt{}/\\
$\text{Ge(Sn,Pb)Bi}_2\text{Te}_4$/\mbt{} family~\cite{yang2025septuple}, which are of interest for realizing QAHE and AHE via interlayer FM coupling. The dispersions in Fig.~\ref{fig4a} demonstrate the formation of a bulk electronic structure characteristic of the WSM phase, with band crossings along the \zgz{} direction and the appearance of Weyl points, similar to the FM phase. As shown in~\cite{shikin2025mgbt}, at a Ge concentration of 25\%, the Weyl phase does not yet form; it is observed only at 37.5\%.

        Modulation of the magnitude of $\lsoc$ (within small limits relative to
        the initial value, taken as 100\%), shown in Figs.~\ref{fig4a}(a1--a5),
        were used to optimize the parameters of the forming Weyl phase. The
        maximum distance between the Weyl points ($0.02\times2 =
        0.04$~\AA$^{-1}$) occurs at $\lsoc = 1.0 \ldots 1.025$. Confirmation
        that this dispersion corresponds to the Weyl state is given in
        Fig.~\ref{fig4a}(a6) (adapted from~\cite{shikin2025mgbt}), which shows
        that the branches forming the Weyl points arise from crossings of CB and
        VB branches, resulting in Weyl nodes with opposite chiralities in the
        \zg{} and \gz{} directions.

        The presented dispersions (Figs.~\ref{fig4a}(a1--a5)) demonstrate
        features  characteristic of a WSM phase (compare with Fig.~\ref{fig3})
        which can confirm our hypothesis of the increasing role of local FM
        interactions as Mn/Ge substitution increases, and the possibility of WSM
        phase formation.
        
It is instructive to compare this with the case of 50\% Ge concentration, for which the dispersions along the \kgz{} direction are shown in Fig.~\ref{fig4a}(a7). One might expect that this configuration would lead to maximal suppression of interlayer AFM Mn\(\downarrow\)/Mn\(\uparrow\) interactions. However, the calculations show that at this level of Mn/Ge substitution, the system no longer exhibits features characteristic of the Weyl phase. In particular, there are no crossings between CB and VB states, which is a necessary condition for the formation of Weyl points with opposite chirality. Hence, the WSM phase is not realized in this case.

It can be assumed that full Mn/Ge substitution in neighboring SLs (i.e., 50\%) completely disrupts the interlayer magnetic coupling. As a result, the local FM-like Mn\(\downarrow\)/Ge/Mn\(\downarrow\) motif, which coexists with the background AFM Mn\(\downarrow\)/Mn\(\uparrow\)/Mn\(\downarrow\) structure, is no longer supported. The system instead resembles MnBi\(_4\)Te\(_7\), where AFM interactions between Mn layers are weakened due to separation by a \bite{} layer~\cite{klimovskikh2020m147}.

        \begin{figure}[H]
            \centering\includegraphics[width=\linewidth]{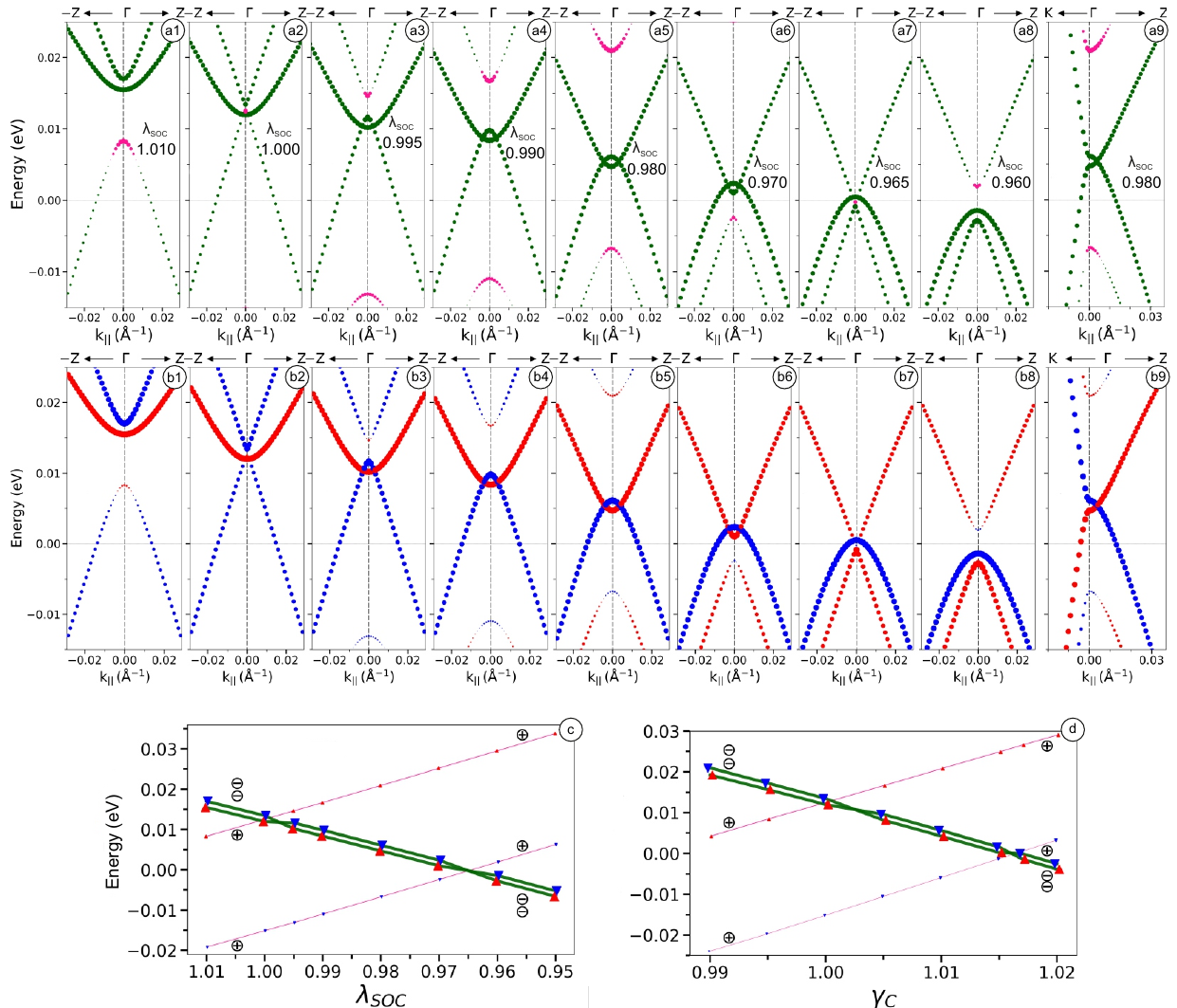}
            
            \caption{(a1--a8) --- Band structures of AFM \mgbtthreeeights{} (\pconfig)
            with $\lsoc$ variation along the \zgz{} direction and (a9)
            corresponding  band structure along the \kgz{} direction for $\lsoc
            = 0.98$ where Weyl node separation is the largest. (b1--b8,~b9) ---
            Detailed view of spin structure evolution corresponding to panels
            (a1--a8,~a9), respectively, where red and blue symbols correspond
            to opposite spin orientations. (c,~d) --- Energy level and parity
            diagrams for $\Gamma$-point eigenstates with dominant Te~$p_z$
            (green) or Bi~$p_z$ (pink) contributions as well as opposite spin
            directions in red and blue for $\lsoc$ and $\gamma_c$ variation,
            respectively. Line thicknesses correspond to magnitude of $|\text{Te
            } p_z - \text{Bi } p_z|$ difference.}


            \label{fig4rest}
        \end{figure}
        
        At the same time, the formation of features in the electronic structure
        characteristic of the Weyl phase is also possible in systems with more
        realistic bulk arrangements of Mn/Ge substitutions, provided that these
        substitutions are uncompensated so that there is a  nonzero average
        total magnetization for any pair of adjacent magnetic layers. This also
        leads to disruption of local interlayer AFM interactions. One example of
        such a system is a $2\times2\times2$ supercell with a 37.5\% Mn/Ge
        substitution concentration and an asymmetric substitution configuration,
        where neighboring SLs differ in substitution concentrations.
        
        Such inequivalent arrangements of Mn/Ge substitution sites may locally
        occur in real bulk crystals in the region of similar concentrations. 
        Figs.~\ref{fig4rest}(a1--a9) show the changes in
        bulk band structure of AFM \mgbtthreeeights{} in the \zgz{} direction in
        \pconfig{} where $\lsoc$ was varied to find largest Weyl point
        separation. This effective SOC may be lowered in practice by additional
        Te/Se substitution for the most optimal Weyl phase since Se has lower
        atomic number.
        
        The changes in the contributions of the Te~$p_z$ and Bi~$p_z$ states are also
        shown here, which correlate with the corresponding representations
        characteristic of the WSM phase in Fig.~\ref{fig3}.
        Fig.~\ref{fig4rest}(a9) shows the band structure along the
        \kgz{} direction, demonstrating that the Weyl points in this system
        arise from crossings between VB and CB branches.
        
        Figs.~\ref{fig4rest}(b1--b9) present corresponding changes in the spin
        structure, showing that the branches forming the Weyl points indeed have
        opposite spin orientations, which is a necessary condition for the WSM
        phase. These results show that a WSM phase with band crossings in the
        \zgz{} direction and the formation of Weyl points can indeed be realized
        in this system. Similar changes in the band and spin structures are also
        observed for \xconfig{} of Mn/Ge substitution sites, see Fig.~5S Suppl.
        
        These calculations demonstrate that a transition to the WSM phase can be
        realized in \mgbt{} TIs with initially AFM interlayer interaction, upon
        increasing Ge concentration, due to inequivalent (uncompensated)
        distribution of Mn/Ge substitution sites in neighboring SLs. In this
        case, the interlayer Mn$\downarrow$/Mn$\uparrow$/Mn$\downarrow$
        interaction transforms into some local FM-like
        Mn$\downarrow$/Ge/Mn$\downarrow$ interactions. It is expected that the
        maximum disruption of AFM interlayer Mn interactions, leading to
        formation of local FM-like Mn$\downarrow$/Ge/Mn$\downarrow$
        interactions, occurs at Ge concentrations of 35--45\%. In this
        concentration range, experimental systems show phases with a minimal
        (near-zero) bulk band gap, which is a necessary condition for the
        formation of a Weyl phase.

    \subsection{Optimization of Weyl phase parameters and analysis of magnetic moment effects}

A recent study~\cite{belopolski2025weylferromagnet} has experimentally demonstrated the formation of a WSM phase in systems based on magnetic FM TIs. In particular, Cr-doped \bite{} with FM interlayer coupling was shown to exhibit a WSM state, characterized by a large Weyl point separation that depends on Cr concentration. This material displays an enhanced anomalous Hall effect (AHE), with conductivity 
\(\sigma_{xy}^\mathrm{3D} = \frac{e^2}{h} \frac{\Delta k_W}{2\pi}\) 
proportional to the Weyl point separation \(\Delta k_W\), a large Hall angle, strong negative magnetoresistance (a hallmark of WSMs), and other distinctive features.

These properties make such magnetic WSM systems highly promising for spintronic applications—potentially surpassing conventional topological insulators in efficiency—and motivate further investigation and optimization of TI-based magnetic WSMs, as well as the search for new materials in this class.

        \begin{figure}[ht!]
            \centering\includegraphics[width=12cm]{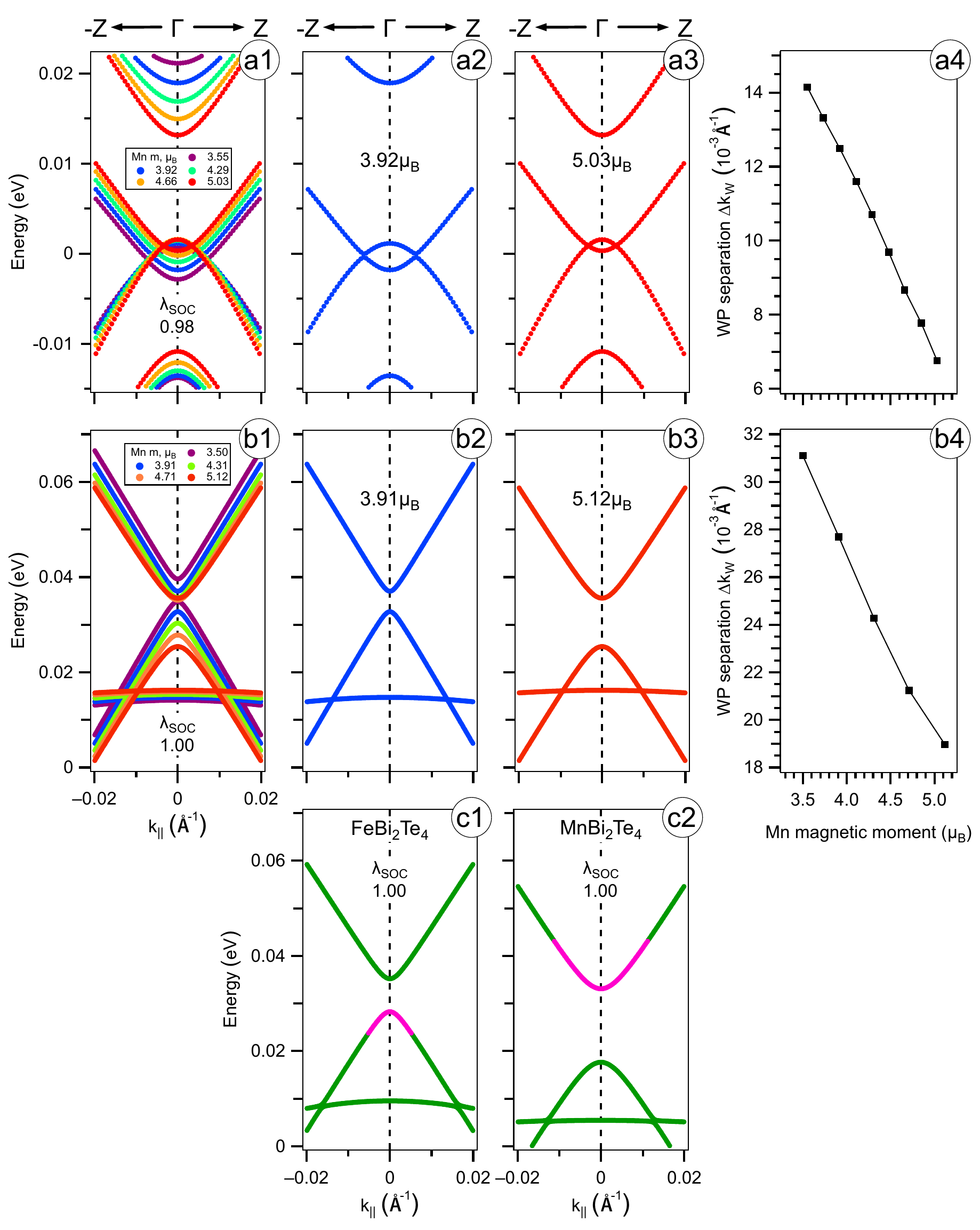}

            \caption{Bulk band structures for various magnetic moments between
            $3.5\mub$ and $5.1\mub$ of (a1) AFM \mgbt{} in \pconfig{} and (b1)
            FM \mbt{}, both are in the WSM phase. Corresponding structures
            with magnetic moments close to typical of Fe ($3.8\mub$) and Mn
            ($5.0\mub$) are shown in panels (a2,~b2) and (a3,~b3), respectively.
            Weyl point separation $\Delta k_W$ dependencies on Mn magnetic
            moment for these systems are shown in (a4,~b4), respectively. Band
            structures of FM \fbt{} (c1) and FM \mbt{} (c2) with unvaried
            magnetic moments $3.8\mub$ and $5.0\mub$,
            respectively, are shown for comparison.}

            \label{fig5}
        \end{figure}

        To understand the role which magnetism plays in bulk band structure of
        WSM and especially how the Weyl point separation $\Delta k_W$ depends on
        it, theoretical calculations were performed for AFM \mgbtthreeeights{}
        and FM \mbt{} with varying Mn magnetic moments from 3.5$\mub$ to
        5.1$\mub$ along the \zgz{} direction. 
        
        The results for these systems are presented in Fig.~\ref{fig5}(a1--a4)
        and Fig.~\ref{fig5}(b1--b4), respectively. Figs.~\ref{fig5}(a1,~b1) show
        general pictures of band structure evolution for these two systems.
        Figs.~\ref{fig5}(a2,~b2) and Fig.~\ref{fig5}(a3,~b3) show band
        structures with magnetic moments of around $3.8\mub$ and $5.0\mub$
        typical of Fe ($3d^6 4s^2$) and Mn ($3d^5 4s^2$), respectively.
        Figs.~\ref{fig5}(a4,~b4) show Weyl point separation $\Delta k_W$
        dependencies on Mn magnetic moments in both systems.
        
It is evident that a reduction in magnetic moment leads to an increase in \(\Delta k_W\) in both AFM \mgbtthreeeights{} and FM \mbt{}. However, for comparable Mn magnetic moments, \(\Delta k_W\) is nearly twice as large in FM \mbt{} (see Fig.~\ref{fig5}(a4,~b4)). This suggests that substituting Mn with Fe may enhance the corresponding WSM phases.

The underlying mechanisms for this behavior differ between the two systems. In FM \mbt{}, the Weyl point separation is governed by the crossing between cone-like and flat-like branches: as the magnetic moment decreases, the cone-like branch shifts upward in energy, while the flat-like branch remains largely unchanged. In contrast, in AFM \mgbt{}, a decrease in magnetic moment increases the energy separation between two valence and two conduction band branches. This causes the valence and conduction states responsible for forming the Weyl points to move toward each other, resulting in an increase in \(\Delta k_W\).

Thus, while both systems exhibit qualitatively similar \(\Delta k_W\) dependence on magnetic moment, the mechanisms driving the band structure evolution are fundamentally different.

        Direct substitution of Mn with Fe in FM \mbt{} yields the FM \fbt{}
        system which shows very similar band structure to that of FM \mbt{} (see
        Fig.~\ref{fig5}(c1,~c2)) where it can also be seen that $\Delta k_W$ is
        greater for FM \fbt{} ($\approx 0.04$~\AA$^{-1}$) than for FM \mbt{}
        ($\approx 0.03$~\AA$^{-1}$).

\begin{figure}[ht!]
            \centering\includegraphics[width=13cm]{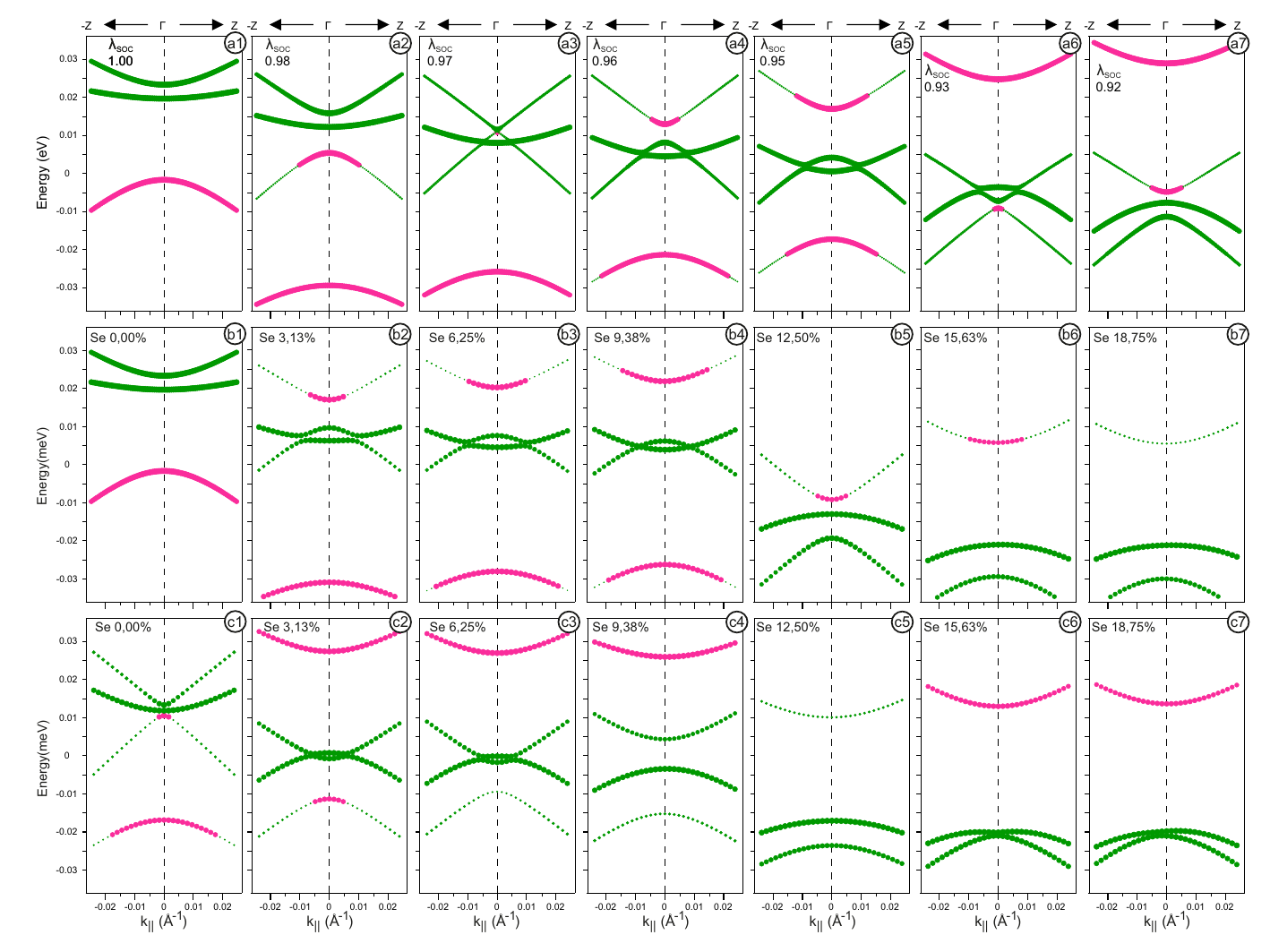}

            \caption{Bulk band structure calculations along the \zgz{} direction
            for  (a1--a7) AFM \fgbtthreeeights{} in \pconfig{} with $\lsoc$
            variation,  (b1--b7) AFM \fgbtthreeeightstese{}, (c1--c7) AFM
            \mgbtthreeeightstese{} for different Te/Se substitution ratios. Green and pink colors
            indicate the dominant contributions from Te~$p_z$ and Bi~$p_z$
            states, respectively.}
            
            \label{fig6}
        \end{figure}

        Fig.~\ref{fig6}(a1--a7) present bulk band structures calculated along
        the \zgz{} direction for AFM \fgbtthreeeights{} system in \pconfig
        (analogous to Fig.~\ref{fig4rest}(a,~b)) for different $\lsoc$ values to find
        largest available WP separation. These results show that formation of
        band structure characteristic of the WSM phase is possible in
        AFM-ordered systems with other magnetic metals (Fe in particular). It
        is also demonstrated that the Weyl point separation $\Delta k_W$ is
        approximately twice as larger in the Fe-based system (see Fig.~\ref{fig6}(a5)) 
        than in the Mn-based system (see Fig.~\ref{fig4rest}(a5)). 

        Large Weyl point separation $\Delta k_W$ is especially critical for
        robust realization of bulk AHE since the latter is directly proportional
        to $\Delta k_W$ (see~\cite{belopolski2025weylferromagnet,
        burkov2014aheweyl, smejkal2018topoafmspintronics} and the formula for
        $\sigma_{xy}^\text{3D}$ above). Notably, for AFM \fgbtthreeeights{}
        shown in Fig.~\ref{fig6}(a) the value of $\Delta k_W$
        is comparable to FM \mbt{} case in Fig.~\ref{fig5}(b3) as well as
        similar WSM phases in the literature~\cite{li2019intrinsic,
        zhang2019topological, li2019controllable, chowdhury2019prediction,
        guo2023novelfbt}.

        In order to find a more realistic system for the formation of the Weyl
        phase (stimulated by SOC strength variation), the calculations of changes
        in the bulk band structure with partial substitution of Te atoms
        by Se atoms were performed. Se atoms have lower atomic number than Te
        atoms, and therefore partial substitution of Te atoms by Se atoms will
        correspond to a decrease in the effective SOC strength with an increase
        in Te/Se substitutions. Fig.~\ref{fig6}(b1--b7) shows the corresponding changes
        in the dispersion dependencies for AFM \fgbtthreeeightstese{} system in \pconfig{}
        in the \zgz{} direction.

        The presented dispersions show that the system actually exhibits band
        structure typical of WSM in Se concentration range from 3\% to 9\%.
        There is an intersection of the corresponding branches of the electronic
        states that form the Weyl points. In this case, the ratio of the
        contributions of Te~$p_z$ and Bi~$p_z$ states has the same character as
        in Fig.~\ref{fig6}(a), Fig.~\ref{fig4a}, Fig.~\ref{fig4rest}. The $\Delta k_W$ values are
        also similar to those shown in Fig.~\ref{fig6}(a). The presented results show
        that, indeed, partial substitution of Te atoms by Se atoms in the
        concentration range of Se atoms from 3\% to 9\% in the AFM \fgbts{}
        system actually leads to the WSM formation and optimization of its
        parameters.

        On the other hand, Fig.~\ref{fig6}(c1--c7) shows, for comparison, the
        corresponding changes in the bulk electronic structure for \mgbtthreeeightstese{} with
        similar Te/Se substitution patterns. In this case, the formation of the
        Weyl phase also occurs, but in a smaller concentration range, from 3\%
        to 6\%. In this case, the formed Weyl phase is not so clearly manifested
        and is characterized by deteriorated parameters.

        From the above one may conclude that the WSM phase in such systems can
        be indeed optimized by selecting appropriate magnetic metal which
        modifies the effective magnetic moment and enlarges the Weyl point
        separation $\Delta k_W$. It can done by substitution of Mn with Fe (in
        combination with the Te/Se substitution). At the same time the variation
        of the SOC strength alone (also by Te/Se substitution) is less effective
        for formation of WSM. These observations further demands more
        detailed investigations.

\section{Conclusions}

We have employed density functional theory to investigate topological phase transitions in \mgbt{} driven by Mn-to-Ge substitution, with a comprehensive analysis of magnetic ordering, spin-orbit coupling (SOC), and uniaxial strain.

In the antiferromagnetic (AFM) phase, increasing Ge concentration induces a transition from a topological insulator (TI) to a normal insulator (NI) via an intermediate Dirac semimetal (DSM) state. This transition is marked by orbital inversion near the \(\Gamma\)-point. In contrast, the ferromagnetic (FM) phase of pristine \mbt{} already hosts a Weyl semimetal (WSM) state, which collapses into a trivial insulator upon Ge doping.

We identified that WSM stability requires band crossings with opposite s$_z$ spin projections along high-symmetry directions. These crossings are suppressed by uniaxial strain or SOC variation (increase and decrease), leading to gap opening and loss of Weyl nodes. Our calculations for the systems with minimal bang gap for representative Ge concentrations (\mgbthalf{}, \mgbtthird{}) and multiple Mn/Ge substitution configurations mapped the critical WSM phase boundaries

Importantly, local asymmetry in Mn/Ge substitution can induce a WSM state even in globally AFM systems, particularly at 37.5\% Ge concentration. This effect arises from the disruption of interlayer AFM stacking and formation of locally FM-like configurations.

We further show that increasing magnetic moment, via Mn\(\rightarrow\)Fe substitution or direct modulation, enlarges the Weyl point separation \(\Delta k_W\), potentially enhancing the anomalous Hall effect. Te\(\rightarrow\)Se substitution provides an additional handle to tune SOC strength and stabilize WSM phases.

Overall, our results establish a robust framework for engineering topological states in magnetic layered compounds and offer practical guidance for experimental realization of tunable Weyl semimetals in the \mbt{} family.

\section{CRediT authorship contribution statement}
A.M. Shikin: Conceptualization, Methodology, Validation, Formal
analysis, Investigation, Writing – original draft, Writing – review \&
editing, Visualization, Project administration, Funding acquisition. 
N.L. Zaytsev: Methodology, Validation, Formal analysis, Investigation,
Data curation, Visualization.
A.V. Eryzhenkov: Methodology, Validation, Formal analysis,
Investigation, Data curation, Writing – review \& editing, Visualization,
Software.
R.V. Makeev: Methodology, Validation, Formal analysis, Investigation,
Data curation, Visualization.
T.P. Makarova: Investigation, Data curation, Visualization. 
D.A. Estyunin: Conceptualization.
A.V. Tarasov: Conceptualization,
Methodology, Validation, Formal analysis, Investigation, Data curation,
Writing – original draft, Writing – review \& editing, Visualization,
Software, Supervision.

\section{Declaration of competing interest}

The authors declare that they have no known competing financial interests or personal relationships that could have appeared to influence the work reported in this paper. 

    \section{Acknowledgements}

        This work was supported by the Russian Science Foundation (Grant
        No.~23-12-00016) and by Saint Petersburg State University (Project
        No.~125022702939-2).

\section{Data Availability}

Data will be made available on request.

\end{document}